\documentclass[amsmath,amssymb,showkeys,superscriptaddress,floatfix,prd,10pt]{revtex4}
\usepackage{graphicx,epstopdf}
\usepackage{subfig}
\usepackage{xcolor}
\usepackage[colorlinks=true, urlcolor=blue, linkcolor=blue, citecolor=green]{hyperref}
\def\pslash{p\!\!\!\slash }
\def\qslash{q\!\!\!\slash }
\def\xslash{x\!\!\!\slash }
\def\eslash{\varepsilon\!\!\!\slash }

\def\vel{\left|}
\def\ver{\right|}

\begin{document}

\title{Electromagnetic multipole moments of the $P_c^+(4380)$ pentaquark in light-cone QCD}

\author{U.~\"{O}zdem}%
\email[]{uozdem@dogus.edu.tr}
\affiliation{Department of Physics, Dogus University, Acibadem-Kadikoy, 34722 
Istanbul, Turkey}
\author{K.~Azizi}%
\email[]{kazizi@dogus.edu.tr}
\affiliation{Department of Physics, Dogus University, Acibadem-Kadikoy, 34722 
Istanbul, Turkey}
\affiliation{School of Physics, Institute for Research in Fundamental Sciences (IPM),
P.~O.~Box 19395-5531, Tehran, Iran}

\date{\today}
 
\begin{abstract}
We calculate the electromagnetic multipole moments
 of the $P_c^+(4380)$ pentaquark by modeling it as the
diquark-diquark-antiquark and $\bar D^*\Sigma_c$ molecular state with quantum numbers $J^P = \frac{3}{2}^-$. 
In particular, the magnetic dipole, electric quadrupole 
and magnetic octupole moments of this particle are extracted in the 
framework of light-cone QCD sum rule.
The values of the  electromagnetic multipole moments obtained
via two pictures differ substantially from each other, which can be used to pin
down the underlying structure of $P_c^+(4380)$.
The comparison of any future experimental data 
on the electromagnetic multipole moments 
of the $P_c^+(4380)$ pentaquark with the 
results of the present work 
 can shed light on the nature and inner quark organization
 of this state.
\end{abstract}

\maketitle

\section{Introduction}
Since the discovery of the X(3872), many charmonium/bottomonium-like
XYZ states have been reported in the experiment.
Some of these hadrons were suggested to have internal structures
more complex than the simple $\bar qq$ configuration for mesons or 
qqq/$\bar q\bar q\bar q$ configuration for baryon/antibaryons in the
conventional picture of the naive quark model, and they are good candidates of exotic hadrons.
In the newly observed family of XYZ, there are some decay channels that break the isospin symmetry 
and affect the identification of the traditional charmonium/bottomonium states negatively. 
The investigation of the properties of these states is one of
the most attractive and active branches of hadron physics.
For some reviews on the 
 theoretical and experimental progress
 on the properties of these new states see Refs.~\cite{Nielsen:2009uh,Swanson:2006st,Voloshin:2007dx,Klempt:2007cp,Godfrey:2008nc,
Faccini:2012pj,Esposito:2014rxa,Chen:2016qju,Ali:2017jda,Esposito:2016noz,Olsen:2017bmm,Lebed:2016hpi}.
In 2015, the LHCb Collaboration discovered 
two candidates of the hidden-charm pentaquark states, $P_c^+(4380)$ and $P_c^+(4450)$,  
in the invariant mass spectrum of $J/\psi\,p$ in the $\Lambda_b^0 \rightarrow J/\psi\,K^-\,p$ decay~\cite{Aaij:2015tga}.
According to the LHCb measurements the $P_c^+(4380)$ has a mass of $4380 \pm 8 \pm 29$~MeV and a width of
$205 \pm 18 \pm 86$~MeV, while the $P_c^+(4450)$ has a mass of $4449.8 \pm 1.7 \pm 2.5$~MeV and a width  of
 $39 \pm 5 \pm 19$~MeV. The preferred spin-parity assignments of the $P_c(4380)$ and $P_c(4450)$ are
 $J^P =3/2^-$ and $5/2^+$, respectively. The minimal quark content of the 
 pentaquarks is $c\bar cuud$  because these states decay into $J/\psi\,p$, and 
 hence they are good candidates of exotic hidden-charm pentaquarks.
After the discovery of LHCb Collaboration there have been intensive theoretical studies to 
explain the properties of these states.
The spectroscopic parameters and decays of the $P_c^+(4380)$ and $P_c^+(4450)$ pentaquarks 
have been studied 
with different models and approaches~\cite{Yang:2015bmv,Burns:2015dwa,Lu:2016nnt,Wang:2016dzu,Shimizu:2017xrg,Shen:2016tzq,Azizi:2016dhy,Roca:2015dva,Chen:2015loa,Huang:2015uda,Meissner:2015mza,Xiao:2015fia,
He:2015cea,Chen:2015moa,Wang:2015qlf,Chen:2016heh,Yamaguchi:2016ote,He:2016pfa,Zhu:2015bba,Lebed:2015tna,Anisovich:2015cia,Maiani:2015vwa,
Ghosh:2015ksa,Wang:2015epa,Scoccola:2015nia,Liu:2017xzo,Guo:2015umn,Liu:2015fea,Guo:2016bkl,Bayar:2016ftu,Lin:2017mtz,Eides:2017xnt,
Azizi:2018bdv,Park:2017jbn,Ortega:2016syt,Shimizu:2016rrd,Chen:2016otp}.
 Different theoretical models give consistent mass results with the experimental observations.
 Hence, more spectroscopic and decay parameters are needed to be calculated and 
 compared with the experimental data. In~\cite{Azizi:2018bdv} it is shown that the molecular picture of $\bar D^*\Sigma_c$
  for $P_c^+(4380)$ gives consistent results for both the mass and width with the experimental data.

As we mentioned above, chasing the announcement of the observation of pentaquarks there have been extensive
amount of studies on their features. However to acquire a deep understanding on their inner structure, which
are still not precise yet, we are in need of more experimental and theoretical studies which may shed light on
their features.
In order to understand the internal structure of the hadrons in the nonperturbative regime
of QCD, the essential challenges are the specification of
the dynamical and statical properties of hadrons 
such as their electromagnetic multipole moments, coupling constants, masses and so on, both theoretically and experimentally. 
Many theoretical models precisely predict the mass and decay 
width of the multiquark states, but the internal structure of these states is still uncertain.
In other words, the mass and decay width alone can not distinguish the internal structure of the multiquark states.
Remember that the electromagnetic multipole moments are equally significant dynamical observables of the multiquark states. 
The electromagnetic multipole moments are directly related
with the charge and current distributions in the hadrons 
and these parameters are directly connected to the spatial
distributions of quarks and gluons inside the hadrons.
Their magnitude and sign provide important information on structure, size and shape of hadrons. 
There are many studies in the literature committed to the 
study the electromagnetic
multipole moments of the standard hadrons,
but unfortunately relatively little are known about the
 electromagnetic multipole moments of the exotic hadrons.
There are a few studies in the literature where the magnetic dipole moment
of the pentaquarks are studied~\cite{Kim:2003ay,Huang:2003bu,Liu:2003ab,Wang:2005gv,Wang:2005jea,Wang:2016dzu,
Bijker:2004gr,Li:2003cb}.

In this study, the magnetic dipole, electric quadrupole and magnetic octupole moments of the 
pentaquark state $P_c^+(4380)$ (hereafter we will denote this state as $P_c$) is extracted by 
using the diquark-diquark-antiquark and $\bar D^*\Sigma_c$ molecular interpolating currents 
in the framework of the light cone QCD sum rule (LCSR). 
The LCSR has already been successfully applied to extract properties of hadrons for decades such as, form factors, 
coupling constants and the electromagnetic multipole
moments. In this approach, the properties of the hadrons are expressed in terms of the light-cone distribution amplitudes (DAs)
and the vacuum condensates~[for details, see for instance~\cite{Chernyak:1990ag, Braun:1988qv, Balitsky:1989ry}].
 Since the electromagnetic multipole moments are expressed in terms of the features of the DAs and the
QCD vacuum, any uncertainty in these parameters reflects the uncertainty of the estimations of
the electromagnetic multipole moments.

The rest of the paper is organized as follows: In section II, the calculation of the sum rules in LCSR will
be presented. In the last section, we numerically analyze the 
sum rules obtained for the electromagnetic multipole moments and discuss
the obtained results.
The explicit expressions of the electromagnetic form factors defining the 
magnetic dipole, electric quadrupole and magnetic octupole moments are moved to the  Appendix A.

\section{The electromagnetic multipole moments of \texorpdfstring{$P_c$}{} pentaquark in LCSR}

In this section we derive the LCSR for the magnetic dipole, electric quadrupole and magnetic octupole moments of the 
 $P_c$ pentaquark. For this purpose, we consider a correlation function 
in the presence of the external electromagnetic field ($\gamma$),
\begin{equation}
 \label{edmn01}
\Pi _{\mu \nu }(q)=i\int d^{4}xe^{ip\cdot x}\langle 0|\mathcal{T}\{J_{\mu}(x)
\bar J_{\nu }(0)\}|0\rangle_{\gamma}, 
\end{equation}%
where $J_{\mu}$ is the interpolating current of $P_c$ pentaquark. In the diquark-diquark-antiquark and molecular pictures, 
it is given as~\cite{Wang:2015epa,Chen:2015moa}
\begin{eqnarray}
 J_{\mu}^{Di}(x)&=&\varepsilon^{abc}\varepsilon^{ade}\varepsilon^{bfg}\big[  u^T_d(x) C\gamma_5 d_e(x)\,u^T_f(x) 
 C\gamma_\mu c_g(x)\, C\bar{c}^{T}_{c}(x)\big],\nonumber\\
J_{\mu}^{Mol}(x)&=&\big[\bar{c}_{d}(x)\gamma_{\mu}d_{d}(x)\big]
\big[\epsilon_{abc}(u_{a}^{T}(x)C\gamma_{\alpha}u_{b}(x))\gamma^{\alpha}\gamma_{5}c_{c}(x)\big],
 \label{eq:JJPc}
\end{eqnarray}
where $C$ is the charge conjugation matrix; and $a$, $b$... are color indices.

The correlation function, given in Eq. (\ref{edmn01}), can be obtained 
in terms of hadronic parameters, known as hadronic representation. 
Furthermore it can be obtained in terms of the quark-gluon parameters 
and distribution amplitudes (DAs) of the photon in the deep Euclidean 
region, known as QCD representation.

The hadronic side of the correlation function
can be obtained  by inserting complete sets of the hadronic pentaquarks,
between the interpolating currents in  Eq. (\ref{edmn01}), with the same
quantum numbers as the ${P_c}$ interpolating currents, i.e.,
\begin{eqnarray}\label{edmn02}
\Pi^{Had}_{\mu\nu}(p,q)&=&\frac{\langle0\mid J_{\mu}\mid
{P_c}(p)\rangle}{[p^{2}-m_{{P_c}}^{2}]}\langle {P_c}(p)\mid
{P_c}(p+q)\rangle_\gamma\frac{\langle {P_c}(p+q)\mid
\bar{J}_{\nu}\mid 0\rangle}{[(p+q)^{2}-m_{{P_c}}^{2}]},
\end{eqnarray}
where q is the momentum of the photon. 
The matrix element of the interpolating current 
between the vacuum and the $P_c$ pentaquark is defined as
\begin{equation}\label{lambdabey}
\langle0\mid J_{\mu}(0)\mid {P_c}(p,s)\rangle=\lambda_{{P_c}}u_{\mu}(p,s),
\end{equation}
where $\lambda_{{P_c}}$ is the  residue and $u_{\mu}(p,s)$ is the
Rarita-Schwinger spinor. 
 Summation over spins of ${P_c}$ pentaquark is applied as:

\begin{align}\label{raritabela}
\sum_{s}u_{\mu}(p,s)\bar u_{\nu}(p,s)=-\Big(\pslash+m_{{P_c}}\Big)\Big[g_{\mu\nu}
-\frac{1}{3}\gamma_{\mu}\gamma_{\nu}-\frac{2\,p_{\mu}p_{\nu}}
{3\,m^{2}_{{P_c}}}+\frac{p_{\mu}\gamma_{\nu}-p_{\nu}\gamma_{\mu}}{3\,m_{{P_c}}}\Big].
\end{align}

The transition matrix element $\langle
{P_c}(p)\mid {P_c}(p+q)\rangle_\gamma$ entering Eq.
(\ref{edmn02}) can be parameterized in terms of four Lorentz invariant form factors as follows
\cite{Weber:1978dh,Nozawa:1990gt,Pascalutsa:2006up,Ramalho:2009vc,Azizi:2008tx,Aliev:2009pd}:
\begin{eqnarray}\label{matelpar}
\langle {P_c}(p)\mid {P_c}(p+q)\rangle_\gamma &=&-e\bar
u_{\mu}(p)\left\{F_{1}(q^2)g_{\mu\nu}\eslash-
\frac{1}{2m_{{P_c}}}\left
[F_{2}(q^2)g_{\mu\nu}+F_{4}(q^2)\frac{q_{\mu}q_{\nu}}{(2m_{{P_c}})^2}\right]\eslash\qslash
\right.\nonumber\\&+&\left.
F_{3}(q^2)\frac{1}{(2m_{{P_c}})^2}q_{\mu}q_{\nu}\eslash\right\} u_{\nu}(p+q),\nonumber\\
\end{eqnarray}
where $\varepsilon$ is the polarization vector of the photon. 

In principle, using the above equations, we can obtain the final expression of the hadronic
side of the correlation function, but we come across with two difficulties: 
all Lorentz structures are not independent and the correlation
function can also receive contributions from spin-1/2 particles, which should be eliminated. 
Actually, the matrix element 
of the current $J_{\mu}$ between vacuum and spin-1/2 pentaquarks is nonzero and is specified as
\begin{equation}\label{spin12}
\langle0\mid J_{\mu}(0)\mid B(p,s=1/2)\rangle=(A  p_{\mu}+B\gamma_{\mu})u(p,s=1/2).
\end{equation}
As is seen the unwanted spin-1/2 contributions are proportional to $\gamma_\mu$ and $p_\mu$.
 By multiplying both sides with $\gamma^\mu$ and using 
 the condition $\gamma^\mu J_\mu = 0$ one can determine the constant A in terms of B.
To remove the spin-1/2 pollutions and obtain only independent structures in the
correlation function, we apply the  ordering for Dirac
matrices as $\gamma_{\mu}\pslash\eslash\qslash\gamma_{\nu}$ and eliminate terms 
with $\gamma_\mu$ at the beginning, $\gamma_\nu$ at the end and those proportional to $p_\mu$ and 
$p_\nu$~\cite{Belyaev:1982cd}. As a result, using Eqs. (\ref{edmn02})-(\ref{matelpar})
for hadronic side we obtain,
\begin{eqnarray}\label{final phenpart}
\Pi^{Had}_{\mu\nu}(p,q)&=&-\frac{\lambda_{_{{P_c}}}^{2}}{[(p+q)^{2}-m_{_{{P_c}}}^{2}][p^{2}-m_{_{{P_c}}}^{2}]}
\Bigg[  -g_{\mu\nu}\pslash\eslash\qslash \,F_{1}(q^2) 
+m_{{P_c}}g_{\mu\nu}\eslash\qslash\,F_{2}(q^2)+
\frac{F_{3}(q^2)}{4m_{{P_c}}}q_{\mu}q_{\nu}\eslash\qslash\, \nonumber\\&+&
\frac{F_{4}(q^2)}{4m_{{P_c}}^3}(\varepsilon.p)q_{\mu}q_{\nu}\pslash\qslash \,+
\mbox{other independent structures} \Bigg].
\end{eqnarray}

The magnetic dipole, $G_{M}(q^2)$, electric quadrupole, 
$G_{Q}(q^2)$, and magnetic octupole, $G_{O}(q^2)$, form factors are
defined in terms of the form factors $F_{i}(q^2)$ in the following
way
 \cite{Weber:1978dh,Nozawa:1990gt,Pascalutsa:2006up,Ramalho:2009vc,Azizi:2008tx,Aliev:2009pd}:
\begin{eqnarray}
G_{M}(q^2) &=& \left[ F_1(q^2) + F_2(q^2)\right] ( 1+ \frac{4}{5}
\tau ) -\frac{2}{5} \left[ F_3(q^2)  +
F_4(q^2)\right] \tau \left( 1 + \tau \right), \nonumber\\
G_{Q}(q^2) &=& \left[ F_1(q^2) -\tau F_2(q^2) \right]  -
\frac{1}{2}\left[ F_3(q^2) -\tau F_4(q^2)
\right] \left( 1+ \tau \right),  \nonumber \\
 G_{O}(q^2) &=&
\left[ F_1(q^2) + F_2(q^2)\right] -\frac{1}{2} \left[ F_3(q^2)  +
F_4(q^2)\right] \left( 1 + \tau \right),\end{eqnarray}
  where $\tau
= -\frac{q^2}{4m^2_{{P_c}}}$. At $q^2=0$, the multipole form factors
are obtained in terms of the functions $F_i(0)$ as:
\begin{eqnarray}\label{mqo1}
G_{M}(0)&=&F_{1}(0)+F_{2}(0),\nonumber\\
G_{Q}(0)&=&F_{1}(0)-\frac{1}{2}F_{3}(0),\nonumber\\
G_{O}(0)&=&F_{1}(0)+F_{2}(0)-\frac{1}{2}[F_{3}(0)+F_{4}(0)].
\end{eqnarray}
The  magnetic dipole ($\mu_{{P_c}}$), electric quadrupole
($Q_{{P_c}}$)  and magnetic octupole ($O_{{P_c}}$) moments are defined in the following way:
 \begin{eqnarray}\label{mqo2}
\mu_{{P_c}}&=&\frac{e}{2m_{{P_c}}}G_{M}(0),\nonumber\\
Q_{{P_c}}&=&\frac{e}{m_{{P_c}}^2}G_{Q}(0),\nonumber\\
O_{{P_c}}&=&\frac{e}{2m_{{P_c}}^3}G_{O}(0).
\end{eqnarray}

The next step is to calculate the correlation function in  Eq.~(\ref{edmn01}) in terms of quark-gluon
parameters as well as the photon DAs in the deep Euclidean region. For this purpose, the interpolating currents are inserted 
into the correlation function and after the contracting out the quark pairs using Wick
theorem the following results are obtained:
\begin{eqnarray}
\label{edmn11}
\Pi^{QCD}_{\mu\nu}(p)&=&i\,\varepsilon^{abc}\varepsilon^{a^{\prime}b^{\prime}c^{\prime}}\varepsilon^{ade}
\varepsilon^{a^{\prime}d^{\prime}e^{\prime}}\varepsilon^{bfg}
\varepsilon^{b^{\prime}f^{\prime}g^{\prime}}\int d^4x e^{ip\cdot x} \langle 0|
\Bigg\{ Tr\Big[\gamma_5 S_d^{ee^\prime}(x) \gamma_5 \tilde S_u^{dd^\prime}(x)\Big]
Tr\Big[\gamma_\mu S_c^{gg^\prime}(x) \gamma_\nu \tilde S_u^{ff^\prime}(x)\Big]\tilde S_c^{c^{\prime}c}(-x) \nonumber\\
&& -  Tr \Big[\gamma_5 S_d^{ee^\prime}(x) \gamma_5 \tilde S_u^{fd^\prime}(x) 
\gamma_\mu S_c^{gg^\prime}(x) \gamma_\nu \tilde S_u^{df^\prime}(x)\Big] \tilde S_c^{c^{\prime}c}(-x)\Bigg \}
|0 \rangle_\gamma ,
\end{eqnarray}
in the diquark-diquark-antidiquark picture, and
\begin{eqnarray}
\Pi_{\mu \nu }^{\mathrm{QCD}}(p)&=&-i\,\epsilon^{abc}\epsilon^{a^{\prime}b^{\prime}c^{\prime}}\,\int d^{4}xe^{ip\cdot x}
\langle 0|
\Bigg\{\mathrm{Tr}\Big[\gamma_{\mu} S_{d}^{dd^{\prime}}(x) 
\gamma_{\nu}S_c^{d^{\prime}d}(-x)\Big] 
\mathrm{Tr}\Big[\gamma_{\beta}\widetilde{S}_u^{aa^{\prime}}(x)\gamma_{\alpha}S_u^{bb^{\prime}}(x)]
\Big( \gamma^{\alpha}\gamma_{5}S_c^{cc^{\prime}}(x)\gamma_{5}\gamma^{\beta}\Big)
\nonumber \\ 
&&-
\mathrm{Tr}\Big[\gamma_{\mu} S_{d}^{dd^{\prime}}(x) 
\gamma_{\nu}S_c^{d^{\prime}d}(-x)\Big]
\mathrm{Tr}\Big[\gamma_{\beta}\widetilde{S}_{u}^{ba^{\prime}}(x)\gamma_{\alpha}S_u^{ab^{\prime}}(x)\Big]
\Big( \gamma^{\alpha}\gamma_{5}S_c^{cc^{\prime}}(x)\gamma_{5}\gamma^{\beta}\Big)\Bigg \}
|0 \rangle_\gamma,
 \label{eq:CorrF2Pc}
\end{eqnarray}%
in the molecular picture, where
\begin{equation*}
\widetilde{S}_{c(q)}^{ij}(x)=CS_{c(q)}^{ij\mathrm{T}}(x)C,
\end{equation*}%
with $S_{q(c)}(x)$ being the quark propagator.
The light (q) and heavy (c) propagators are given as~\cite{Balitsky:1987bk}
\begin{align}
\label{edmn12}
S_{q}(x)=i \frac{{\xslash}}{2\pi ^{2}x^{4}} 
- \frac{ \bar qq }{12} 
- \frac{ \bar q \sigma. G q }{192} x^2
-\frac {i g_s }{32 \pi^2 x^2} G^{\mu \nu}(x) \Big[\rlap/{x} 
\sigma_{\mu \nu} 
+  \sigma_{\mu \nu} \rlap/{x}
 \Big],
\end{align}%
and
\begin{align}
\label{edmn13}
S_{c}(x)&=\frac{m_{c}^{2}}{4 \pi^{2}} \Bigg[ \frac{K_{1}\Big(m_{c}\sqrt{-x^{2}}\Big) }{\sqrt{-x^{2}}}
+i\frac{{\xslash}~K_{2}\Big( m_{c}\sqrt{-x^{2}}\Big)}
{(\sqrt{-x^{2}})^{2}}\Bigg]
-\frac{g_{s}m_{c}}{16\pi ^{2}} \int_0^1 dv\, G^{\mu \nu }(vx)\Bigg[ (\sigma _{\mu \nu }{\xslash}
  +{\xslash}\sigma _{\mu \nu })\frac{K_{1}\Big( m_{c}\sqrt{-x^{2}}\Big) }{\sqrt{-x^{2}}}\nonumber\\
&+2\sigma_{\mu \nu }K_{0}\Big( m_{c}\sqrt{-x^{2}}\Big)\Bigg],
\end{align}%
where $K_i$ are modified the second kind Bessel functions and $G^{\mu\nu}$ is the gluon field strength tensor.
Note that with the above form of the light quark propagator and considering Eqs. (\ref{edmn11}) and (\ref{eq:CorrF2Pc}),
 which represent the quark propagators between vacuum and the photon states, we include 
 all the possible contributions.

The correlation function includes  different types of contributions.
In the first part, the photon interacts with one of the light or heavy quarks, perturbatively. In this case,  
the propagator of the quark that interacts with the photon, perturbatively   is replaced by
\begin{align}
\label{free}
S^{free}(x) \rightarrow \int d^4y\, S^{free} (x-y)\,\rlap/{\!A}(y)\, S^{free} (y)\,,
\end{align}
with $S^{free}(x)$ representing the first term of the light or heavy quark propagator,
and the remaining four propagators in Eqs.~(\ref{edmn11}) and (\ref{eq:CorrF2Pc}) are replaced with the full quark propagators
including the free (perturbative) part as well as the interacting parts (with gluon or QCD vacuum)  as nonperturbative
contributions. The full perturbative contribution  is obtained by applying the  above replacement for the perturbatively interacting quark propagator with the photon and   replacing the remaining propagators  by their free parts.


In the second type, one of the light quark propagators in Eqs.~(\ref{edmn11}) and (\ref{eq:CorrF2Pc}), describing the photon emission at large distances, is replaced by
\begin{align}
\label{edmn14}
S_{\alpha\beta}^{ab}(x) \rightarrow -\frac{1}{4} \big[\bar{q}^a(x) \Gamma_i q^b(x)\big]\big(\Gamma_i\big)_{\alpha\beta},
\end{align}
and the remaining propagators are replaced with the full quark propagators.
 Here, $\Gamma_i$ are the full set of Dirac matrices. Once 
Eq. (\ref{edmn14}) is plugged into Eqs. (\ref{edmn11}) and (\ref{eq:CorrF2Pc}) , there appear matrix
elements such as $\langle \gamma(q)\vel \bar{q}(x) \Gamma_i q(0) \ver 0\rangle$
and $\langle \gamma(q)\vel \bar{q}(x) \Gamma_i G_{\alpha\beta}q(0) \ver 0\rangle$,
representing the nonperturbative contributions. 
These matrix elements can be expressed in terms of photon wave functions with definite
twists, whose expressions are given in Ref.~\cite{Ball:2002ps}.
The QCD side of the correlation function can be obtained in terms of quark-gluon properties 
 using Eqs.~(\ref{edmn11})-(\ref{edmn14}) and after applying the Fourier transformation to 
transfer the calculations to the momentum space.

The two representations, the QCD and hadronic sides, of the correlation function, in two different kinematical regions
are then matched using dispersion relation. Then we carry out the double
Borel transforms with respect to the variables $p^2$ and $(p+q)^2$ on both sides of
the correlation function in order to suppress the contributions of the higher states and continuum, 
and use the quark-hadron duality assumption. 
By matching the coefficients of the structures $g_{\mu\nu}\pslash\eslash\qslash$, $g_{\mu\nu}\eslash\qslash$,
$q_{\mu}q_{\nu}\eslash\qslash$ and  
$(\varepsilon.p)q_{\mu}q_{\nu}\pslash\qslash$, respectively for the $F_1$, $F_2$, $F_3$ and $F_4$  
 we find LCSR for these four invariant form factors. 
The explicit expressions of the sum rules for these form factors are given in
the Appendix A. For the sake of simplicity only the results obtained from the diquark-diquark-antiquark picture are given.
The results of the molecular picture have more or less has similar forms.

\section{Numerical analysis and conclusion}

Present section is devoted to the numerical analysis for the magnetic dipole, electric quadrupole 
and magnetic octupole moments of the $P_c$ pentaquark. 
We use $m_u=m_d=0$, $m_c = (1.28\pm 0.03)\,GeV$~\cite{Patrignani:2016xqp}, 
$m_{P_c}= 4.38\pm 0.37~GeV$~\cite{Patrignani:2016xqp}, 
$f_{3\gamma}=-0.0039~GeV^2$~\cite{Ball:2002ps}, 
$\langle \bar uu\rangle = 
\langle \bar dd\rangle=(-0.24\pm0.01)^3\,GeV^3$ \cite{Ioffe:2005ym},
and 
$\langle g_s^2G^2\rangle = 0.88~ GeV^4$~\cite{Nielsen:2009uh}.
To obtain a numerical prediction for the electromagnetic multipole moments, 
we also need to specify the values of the residue of the $P_c$ pentaquark.
The residue is obtained from the mass sum rule as $\lambda_{P_c}=(1.55 \pm 0.28)\times 10^{-3}~GeV^6$~\cite{Wang:2015epa}
for the diquark-diquark-antiquark picture and $\lambda_{P_c}=(0.98 \pm 0.05)\times 10^{-3}~GeV^6$~\cite{Azizi:2016dhy}
for molecular picture. 
The parameters used in the photon distribution amplitudes are given in \cite{Ball:2002ps}.

The predictions for the magnetic dipole, electric quadrupole 
and magnetic octupole moments depend on two
 auxiliary parameters; the Borel mass parameter $M^2$ and continuum threshold $s_0$. 
 According to the standard prescriptions in the method used the predictions should  
 weakly depend on these helping parameters.
 The continuum threshold represents the scale at which, the excited states and continuum start to
contribute to the correlation function.
To specify the working interval of the continuum
threshold, we impose the conditions of pole dominance
and operator product expansion (OPE) convergence. 
Our numerical computations lead
to the interval [22-24]~$GeV^2$ for this parameter.
To specify the working region of the Borel parameter
one needs to take into account two criteria: convergence
of the series of OPE and effective suppression of the
higher states and continuum .
The above requirements restrict the working
region of the Borel parameter to $5~GeV^2 \leq M^2 \leq 7~GeV^2$. 
 In Fig. 1, we plot the dependencies of the magnetic dipole, electric quadrupole and 
 magnetic octupole moments on $M^2$ 
 at several fixed values of the continuum threshold $s_0$.  
 As can be seen from this figure, the corresponding electromagnetic multipole moments
 show overall weak dependence on the variations of the  Borel mass parameter in its working regions. However, the dependence of the results on the continuum threshold is considerable.

In this part we would like to discuss the the amount of the perturbative and different nonperturbative  contributions to the whole results.
Our numerical calculations show that almost $85\%$ of the total 
contribution belongs to the
perturbative  part and the remaining $15\%$ corresponds to the 
nonperturbative contributions: almost $17\%$ of the total 
nonperturbative contributions comes from the terms containing
quark condensates $\langle \bar qq \rangle$, $5\%$ belongs to those containing gluon
condensates $\langle g_s^2 G^2\rangle$, $77\%$ belongs to the terms including the DAs 
parameters 
and remaining $1\%$ corresponds to the higher dimensional operators, 
where because of their negligible contributions we will not present these terms in the Appendix.

  \begin{figure}
\centering
 \includegraphics[width=0.35\textwidth]{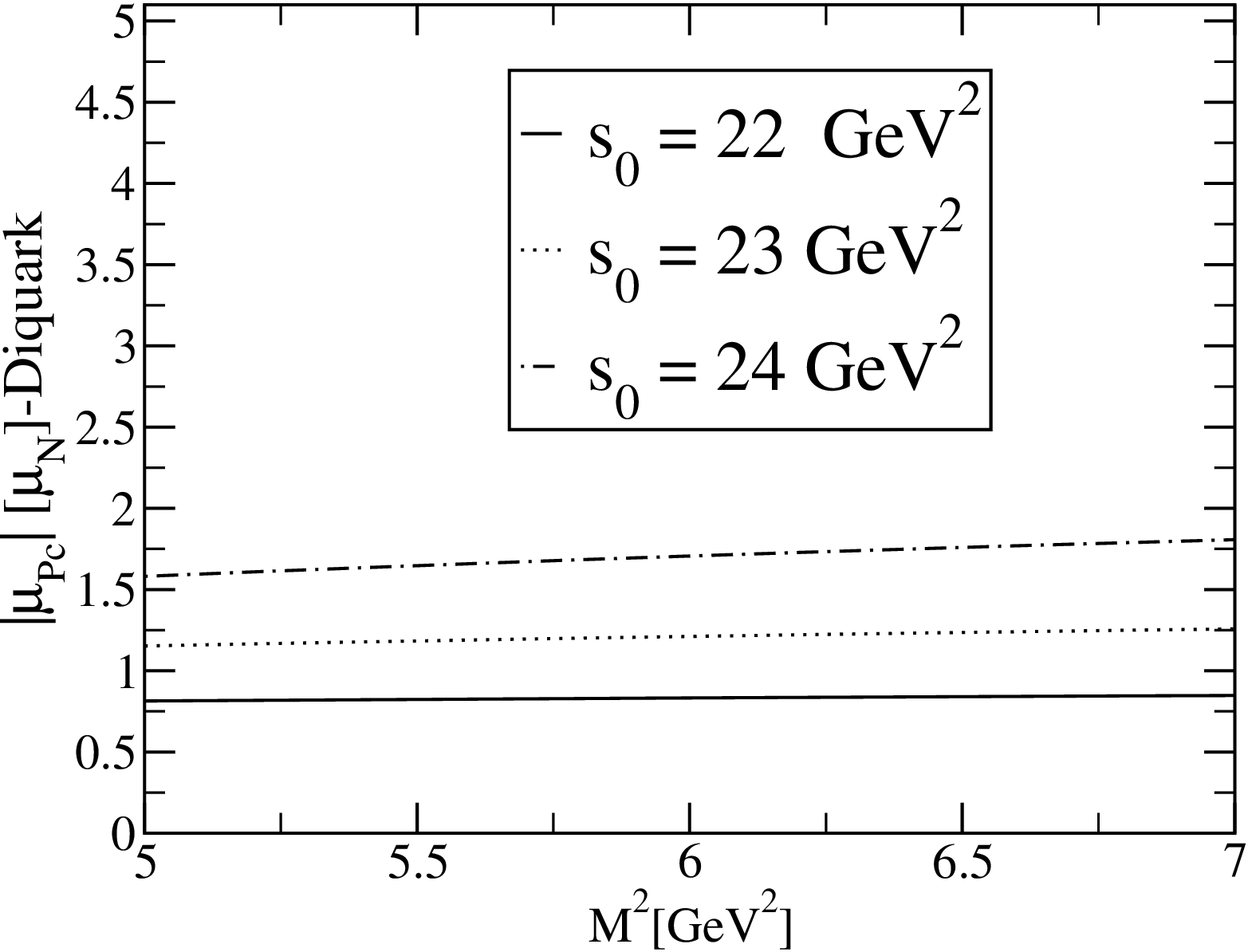}~~~~~~~
  \includegraphics[width=0.35\textwidth]{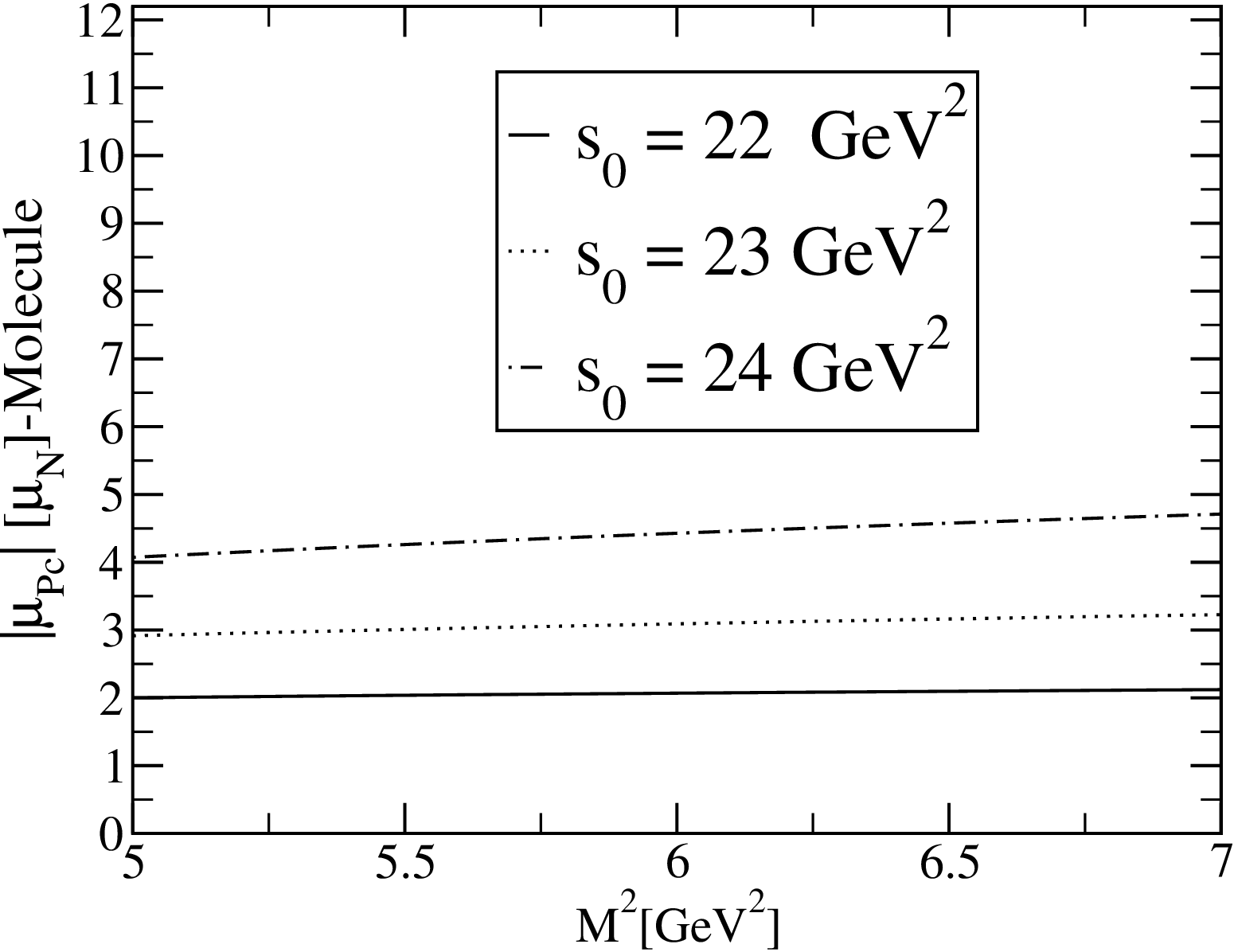}\\
  \vspace{1cm}
 \includegraphics[width=0.35\textwidth]{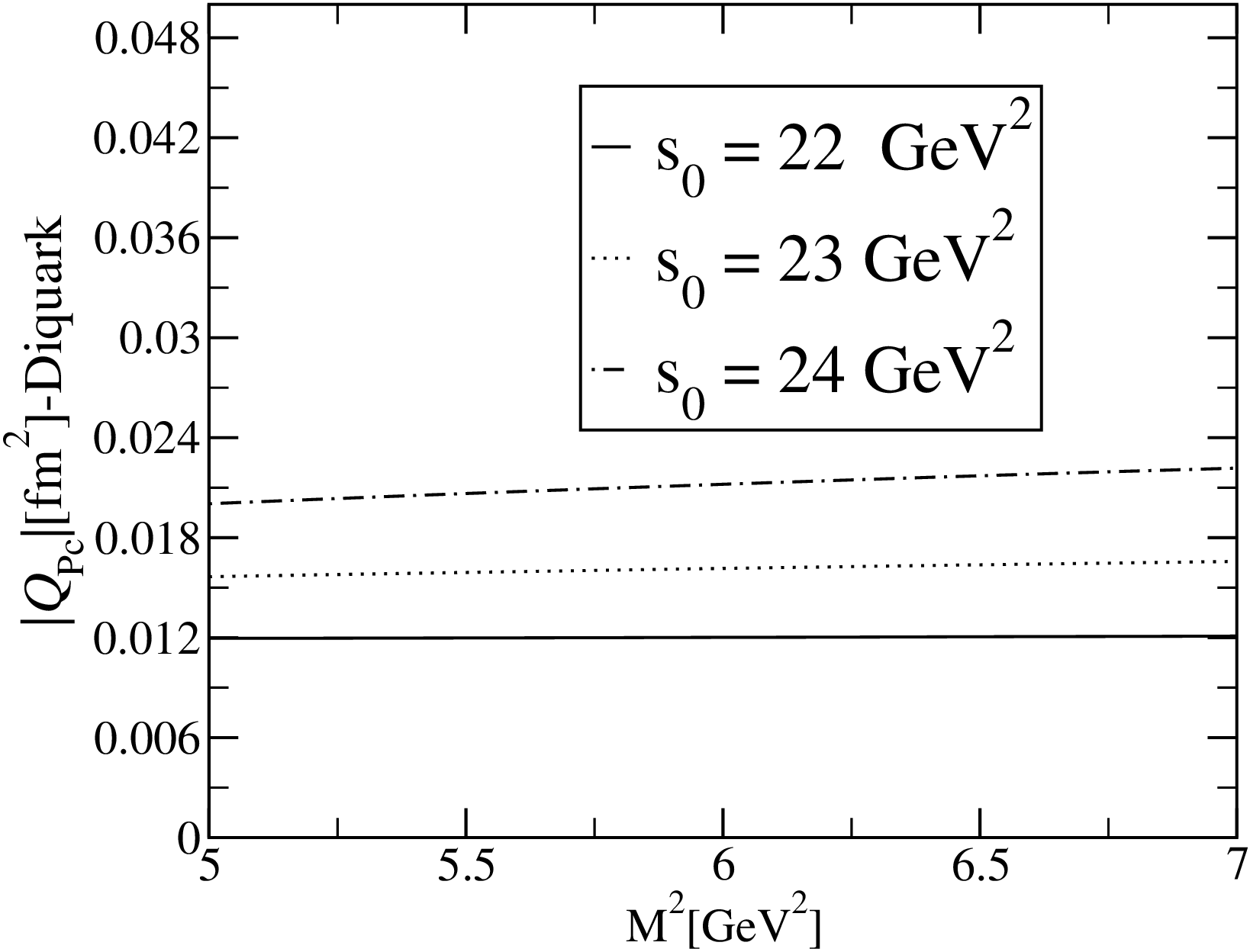}~~~~~~~
  \includegraphics[width=0.35\textwidth]{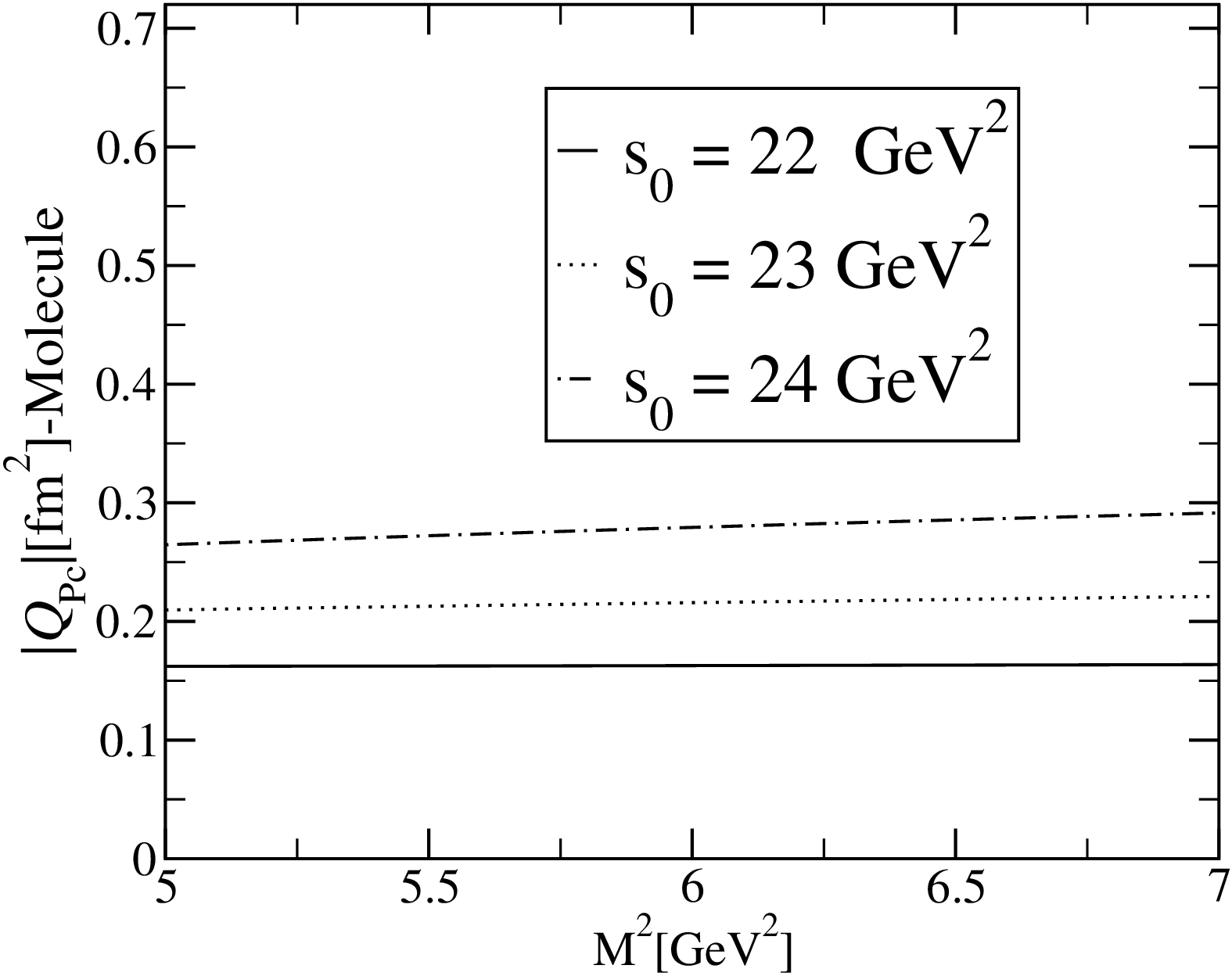}\\
  \vspace{1cm}
 \includegraphics[width=0.35\textwidth]{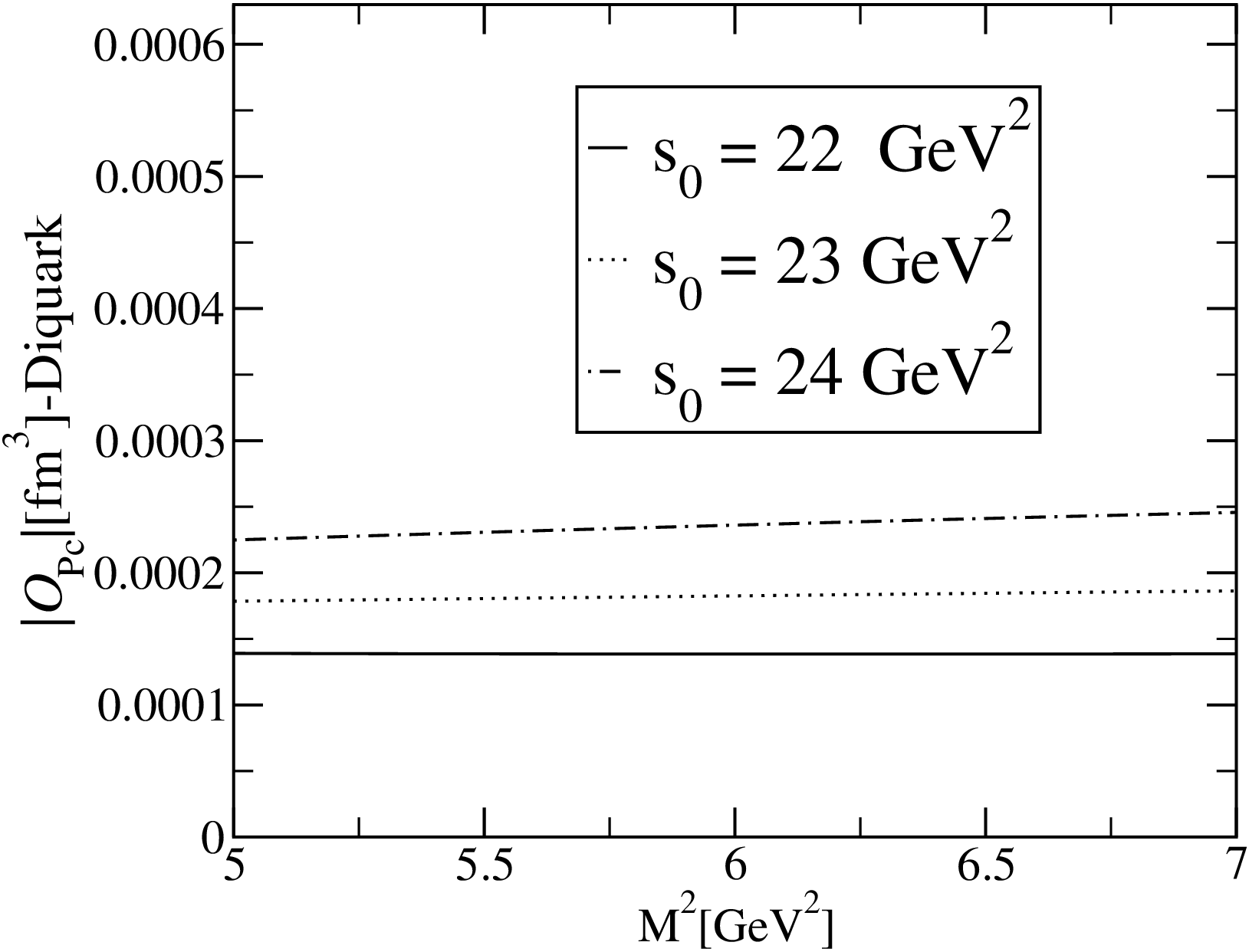}~~~~~~~
  \includegraphics[width=0.35\textwidth]{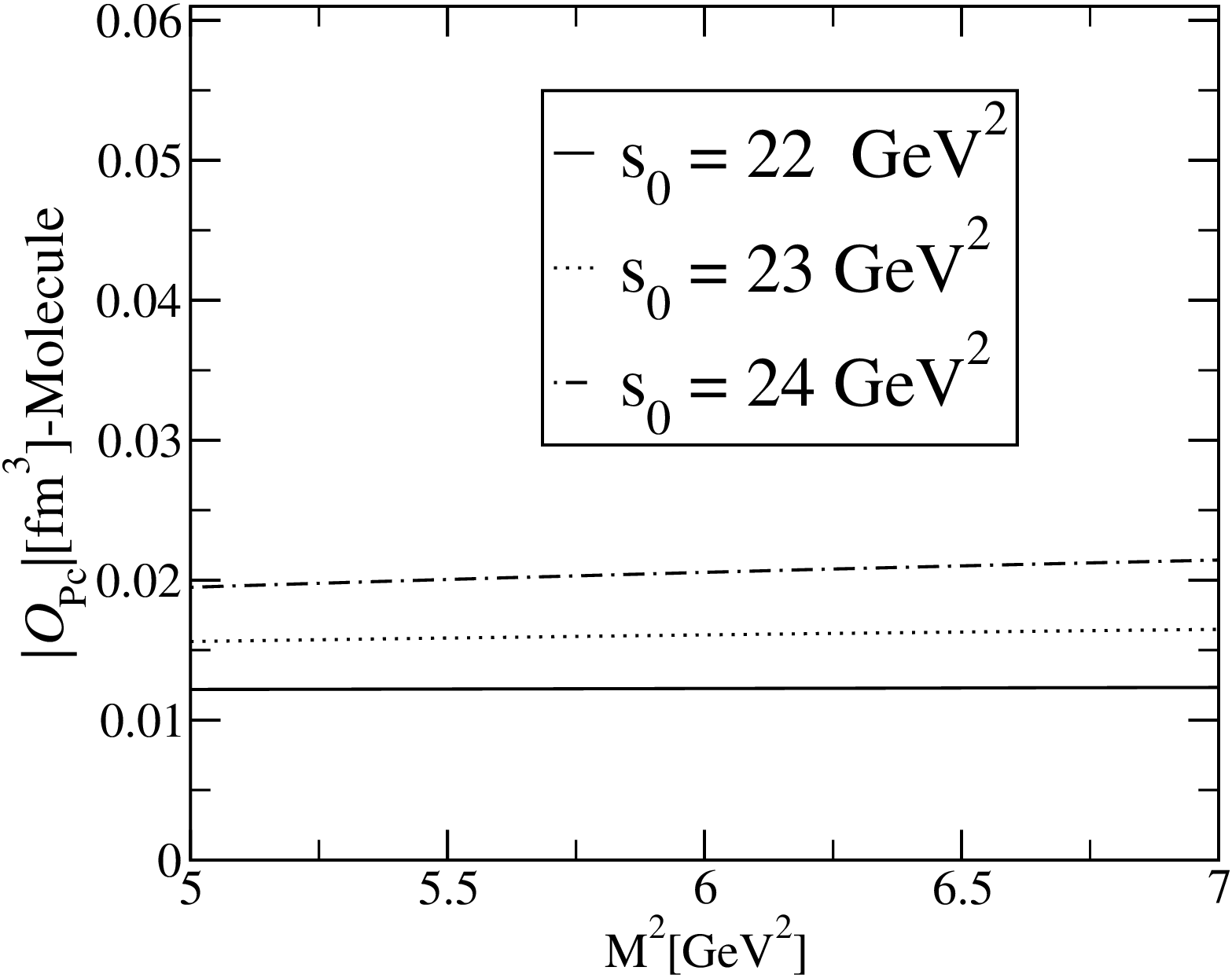}
 \caption{ The dependence of the magnetic dipole, electric quadrupole and magnetic octupole moments for $P_c$ pentaquark;
 on the Borel parameter squared $M^{2}$
 at different fixed values of the continuum threshold.}
  \end{figure}
%

Our final results for the magnetic dipole, electric quadrupole and magnetic octupole moments are given in Table I.
The errors in the given results arise due to the variations in the  calculations of the working regions of
 $M^2$ and $s_0$ as well as the uncertainties 
in the values of the input parameters and the photon DAs. 
We shall remark that the main source of uncertainties is due to the variations of the results  with respect to $s_0$. As previously mentioned, the continuum threshold is not totally arbitrary but it depends on the energy of the first excited state. We don't  have  enough information on the mass of the first excited state in the channel under consideration. Hence we choose its working interval such that the above mentioned criteria of the sum rules be satisfied. Our analyses show that in the selected region for  $s_0$, the dependence of the results on this parameter is very weak compared to the regions out of its working window. 
We also would like to note that in Table I and Fig. 1, the absolute values
are given since it is not possible to specify the sign of the residue from
the mass sum rules. Hence, it is not possible to predict the signs of the magnetic dipole,
 electric quadrupole and magnetic octupole moments.
\begin{table}[t]
	\addtolength{\tabcolsep}{10pt}
	\begin{center}
\begin{tabular}{cccccccccc}
	   \hline\hline
	   Picture&  $|\mu_{P_c}|[\mu_N]$	& $|Q_{P_c}|[fm^2]$		&$|O_{P_c}|[fm^3]$   \\
	   \hline\hline
	   Diquark&   $1.30 \pm 0.50$     &$0.017 \pm 0.05$ 	&$0.0002 \pm 0.00006$\\
	   Molecule&  $3.35 \pm 1.35$     &$0.23 \pm 0.06$ 	&$0.017 \pm 0.004$\\
	   \hline\hline
\end{tabular}
\end{center}
\caption{Results of the magnetic dipole, electric quadrupole and magnetic octupole moments of $P_c$ pentaquark.}
	\label{table}
\end{table}

In conclusion, we have calculated the electromagnetic multipole moments
 of the $P_c^+(4380)$ pentaquark by modeling it as the
diquark-diquark-antiquark and molecular state of $\bar D^* \Sigma_c$ with quantum numbers $J^P = \frac{3}{2}^-$. 
The magnetic dipole, electric quadrupole 
and magnetic octupole moments of this particle have been extracted in the 
framework of light-cone QCD sum rule.
The values of the  electromagnetic multipole moments obtained
via two pictures show large differences from each other, which can be used to pin
down the underlying structure of $P_c^+(4380)$.
In other words, as many models give compatible results on the mass and width with the experimental data
preventing us assigning exact inner structure for pentaquarks,
the experimental measurement of the electromagnetic multipole moments 
of the $P_c^+(4380)$ pentaquark indeed can
help us precisely distinguish its inner structure. 
The electromagnetic multipole moments of $P_c^+(4380)$ can be extracted
 through the process $\gamma^{(*)}p\rightarrow P_c^+(4380) \rightarrow 
P_c^+(4380)\, \gamma \rightarrow  J/\psi\, p\, \gamma $ like those of $\Delta^+$ baryon.

\section{Acknowledgements}

This work has been supported by the Scientific and
Technological Research Council of Turkey (T\"{U}B\.{I}TAK)
under the Grant No. 115F183.
\newpage
\section*{Appendix A: Explicit forms of the sum rules for \texorpdfstring{$F_i^{Di}$}{} }
In this appendix, we present the explicit expressions for the sum rules $F_i^{Di}$: 
\begin{align}
 F_1^{Di}&=-\frac{e^{m^2_{P_c}/M^2}}{\lambda_{P_c}^2}\Bigg\{
 \frac{m_c\,\langle \bar qq \rangle}{3774873\,\pi^5}\Bigg[-40\,\pi^2\,f_{3\gamma}(e_d+10e_u)\,
 I_2[\mathcal{V}]\,I[0,4,4,0]+3(e_d+14e_u)\,I_4[\tilde S]\, I[0,5,4,0]\nonumber\\
 &+192\,e_c\,
 \Big(I[0,5,2,2]-2\,I[0,5,2,3]+I[0,5,2,4]-2\,I[0,5,3,2]+2\,I[0,5,3,3]+I[0,5,4,2]\Big)\Bigg]\nonumber\\
 &-\frac{\langle g_s^2 G^2 \rangle}{108716359680\,\pi^7}\Bigg[20\,\pi^2\,f_{3\gamma}
 \Big(4\,(32e_d+41e_u)\,I[0,4,4,0]+3\,(9e_d+20e_u)\,I[0,4,5,0]\Big)I_2[\mathcal{V}]+(72e_d+720e_u\nonumber\\
 &-702e_c)\,I[0,5,2,1]-(324e_d+2184e_u-2673e_c)\,I[0,5,2,2]+(492e_d+2456e_u-3511e_c)\,I[0,5,2,4]\nonumber\\
 &-(300e_d+1240e_u-1811e_c)\,I[0,5,2,4]+(60e_d+248e_u-271e_c)\,I[0,5,2,5]+(216e_d+2160e_u-2106)\nonumber\\
 &\times I[0,5,3,1]+(684e_d+4728e_d-5693e_c)\,I[0,5,3,2]-(624e_d+3424e_u-4483e_c)\,I[0,5,3,3]+(156e_d\nonumber\\
 &+856e_u-893e_c)\,I[0,5,3,4]+(216e_d+2160e_u-2106e_c)\,I[0,5,4,1]-(396e_d+2904e_u-3375e_c)\nonumber\\
 &\times I[0,5,4,2] +(132e_d+968e_u-973e_c)\,I[0,5,4,3]
 -(72e_d+720e_u-702e_c)\,I[0,5,5,1]+(36e_d+360e_u\nonumber\\
 &-351e_c)\,I[0,5,5,2]\Bigg]\nonumber\\
 &-\frac{f_{3\gamma}}{3019898880\,\pi^5}(13e_d+58e_u)\,I_2[\mathcal{V}]\,I[0,6,5,0]\nonumber\\
 &+\frac{e_c}{880803840\,\pi^7}\Bigg[4\,I[0,7,2,3]-13\,I[0,7,2,4]+15\,I[0,7,2,5]
 -7\,I[0,7,2,6]+I[0,7,2,7]-12\,I[0,7,3,3]\nonumber\\
 &+27\,I[0,7,3,4]-18\,I[0,7,3,5]+3\,I[0,7,3,6]+12\,I[0,7,4,3]+3\,I[0,7,4,5]-4\,I[0,7,5,3]+I[0,7,5,4]\Bigg]\Bigg\},\\
 \nonumber\\
 F_2^{Di}&=\frac{m_{P_c}\,e^{m^2_{P_c}/M^2}}{\lambda_{P_c}^2}\Bigg\{
   -\frac{m_c\,\langle \bar qq \rangle}{125829120\,\pi^5}\Bigg[
 -120\, (e_d + 10 e_u)\, f_{3\gamma}\, \pi^2\, I_2[\mathcal V]\, I[0, 4, 4, 0]
+ (3 e_d + 14 e_u) I_4[\mathcal S]+ 2 e_d\, I_4[\mathcal T_1] \nonumber\\
 &\times I[0, 5, 4, 0]+64\,e_c\,\Big(
  I[0, 5, 2, 2] - 2\,I[0, 5, 2, 3] + I[0, 5, 2, 4] - 
    2\,I[0, 5, 3, 2] + 2\,I[0, 5, 3, 3] + I[0, 5, 4, 2] \Big)\Bigg]\nonumber\\
    &+\frac{\langle g_s^2 G^2 \rangle}{108716359680\,\pi^7}\Bigg[
    20\, f_{3\gamma} \pi^2\, \Big\{-4 \Big((52 e_d - 19 e_u) I_2[\mathcal A] - 
      9 (32 e_d + 41 e_u) I_2[\mathcal V] + 4 (3 e_d + 2 e_u)\, I_6[\psi^\nu]\Big)\,I[0, 4, 4, 0]\nonumber\\
      &+3 \Big(8 (-34 e_d + e_u) I_2[\mathcal A] + 9 (9 e_d + 220 e_u)\, I_2[\mathcal V]\Big)\,I[0, 4, 5, 0]\Big\}
      -\Big\{
      (-72 e_d + 702 e_c - 720 e_u)\,I[0, 5, 2, 1]+ (324 e_d \nonumber \\
    &- 2673 e_c + 2184 e_u)\,I[0, 5, 2,2] 
    + (-492 e_d + 3511 e_c - 2456 e_u)\,I[0, 5, 2, 
   3] + (300 e_d - 1811 e_c + 1240 e_u)\,I[0, 5, 2,4] \nonumber\\
   &+ (-60 e_d + 271 e_c - 248 e_u)\,I[0, 5, 2, 
   5] + (216 e_d - 2106 e_c + 2160 e_u)\,I[0, 5, 3, 
   1] + (-684 e_d + 5697 e_c - 4728 eu)\nonumber\\
   &\times I[0, 5, 3,2] + (624 e_d - 4484 e_c + 3424 e_u)\,I[0, 5, 3, 3] 
   + (-156 e_d + 893 e_c - 856 e_u)\,I[0, 5, 3, 4] \nonumber\\
       +& (-216 e_d + 2106 e_c - 2160 e_u)\,I[0, 5, 4, 1] + (396 e_d - 3375 e_c + 2904 e_u)\,I[0, 5, 4, 
   2] + (-132 e_d + 973 e_c - 968 e_u)\nonumber\\
   & \times I[0, 5, 4, 3] + (72 e_d - 702 e_c + 720 e_u)\,I[0, 5, 5, 
   1] + (-36 e_d + 351 e_c - 360 e_u)\,I[0, 5, 5, 2]
      \Big\}\Bigg]\nonumber\\
   &+\frac{f_{3\gamma}}{3019898880\,\pi^5}    (13 e_d + 58 e_u) I_2[\mathcal V]\,I[0, 6, 5, 0]\nonumber\\
   &-\frac{e_c}{880803840\, \pi^7}\Bigg[
   4\,I[0, 7, 2, 3] - 13\,I[0, 7, 2, 4] + 15\,I[0, 7, 2, 5] - 
 7\,I[0, 7, 2, 6] + I[0, 7, 2, 7] - 12\,I[0, 7, 3, 3]\nonumber\\
 &+ 
 27\,I[0, 7, 3, 4] - 18\,I[0, 7, 3, 5] + 3\,I[0, 7, 3, 6] + 
 12\,I[0, 7, 4, 3] - 15\,I[0, 7, 4, 4] + 3\,I[0, 7, 4, 5] - 
 4\,I[0, 7, 5, 3]\nonumber\\
 &+ I[0, 7, 5, 4]\Bigg]\Bigg\},
\end{align}
\begin{align}
 F_3^{Di}&=\frac{4\,m_{P_c}\,e^{m^2_{P_c}/M^2}}{\lambda_{P_c}^2}\Bigg\{
\frac{m_c\,\langle g_s^2G^2 \rangle}{5435817984\, \pi^7}\Bigg[
 2\,f_{3\gamma}\,\pi^2\, I_2[\mathcal V] \,\big(4\,(5 e_d - 16 e_u)\,I[0, 3, 3, 0] + 
   3\,(3 e_d + 5 e_u)\,I[0, 3, 4, 0]\big)
   +3\, f_{3\gamma}\,\pi^2 \nonumber\\
   &\times I_4[\mathcal V]\, \big(14\,(e_d+ 4 e_u)\,I[0, 3, 3, 0] 
   - (9 e_d + 194 e_u)\,I[0, 3, 4, 0]\big)
   (-36 e_d + 351 e_c - 360 e_u)\,I[0, 4, 1, 2] 
   + (108 e_d \nonumber\\
  &- 977 e_c + 1080 e_u)\, I[0, 4, 1, 3]
   + (-108 e_d + 901 e_c - 1080 e_u)\,I[0, 4, 1, 4] 
   + (36 e_d - 275 e_c + 360 e_u)\,I[0, 4, 1, 5] 
    \nonumber\\
   &+ (108 e_d - 1053 e_c+ 1080 e_u)\,I[0, 4, 2, 2] 
   + (-216 e_d + 1954 e_c - 2160 e_u)\,I[0, 4, 2,3] 
   + (108 e_d - 901 e_c + 1080 e_u) \nonumber\\
   &\times I[0, 4, 2, 4]+ (-108 e_d + 1053 e_c - 1080 e_u)\,I[0, 4, 3, 2] 
   + (108 e_d - 977 e_c + 1080 e_u)\,I[0, 4, 3, 3]  \nonumber\\
   & + (36 e_d - 351 e_c + 360 e_u)\, I[0, 4, 4, 2]
   + (-24 e_d - 240 e_u)\,I[1, 3, 1, 3] + (72 e_d + 720 e_u)\,I[1, 3,1, 4] + (-72 e_d \nonumber\\
   &- 720 e_u)\,I[1, 3, 1, 5] 
   + (24 e_d + 240 e_u) I[1, 3, 1, 6] + (72 e_d + 720 e_u)\,I[1, 3, 2, 3] + (-144 e_d - 1440 e_u)\,I[1, 3, 2, 4] 
    \nonumber\\
   &+ (72 e_d + 720 e_u)\,I[1, 3, 2, 5] + (-72 e_d- 720 e_u)\,I[1, 3, 3, 3] + (72 e_d + 720 e_u)\,I[1, 
   3, 3, 4] + (24 e_d + 240 e_u)\nonumber\\
   &\times I[1, 3, 4, 3]\Bigg]\nonumber\\
   &-\frac{m_c\,f_{3\gamma}}{251658240 \pi^5}\Bigg[(e_d + 4 e_u)\, 
   I_2[\mathcal V] + 6 (3 e_d + 14 e_u)\, I_4[\mathcal V]\Bigg]\, I[0, 5, 4, 0]\nonumber\\
   &+\frac{m_c\,e_c}{31457280 \pi^7}\Bigg[
   I[0, 6, 1, 4] - 3\,I[0, 6, 1, 5] + 3\,I[0, 6, 1, 6] - 
 I[0, 6, 1, 7] - 
 3\,\Big(I[0, 6, 2, 4] - 2\,I[0, 6, 2, 5]  \nonumber\\
 &+ I[0, 6, 2, 6]- I[0, 6, 3, 4] + I[0, 6, 3, 5]\big) - I[0, 6, 4, 4]\Bigg]\Bigg\},
 \end{align}
 and
 \begin{align}
 F_4^{Di}&=\frac{4\,m^3_{P_c}\,e^{m^2_{P_c}/M^2}}{\,\lambda_{P_c}^2}\Bigg\{
 \frac{m_c\, \langle g_s^2G^2\rangle}{452984832 \pi^7}\Bigg[
 3\,f_{3\gamma}\, \pi^2\, I_4[\mathcal V] \Big( (3 e_d - 8 e_u)\,I[0, 2, 3, 0] + 13 e_u \,I[0, 2, 4, 0]\Big)
 +4\,(e_d + 10 e_u)  \nonumber\\
 &\times \Big(I[0, 3, 1, 3]- 3\,I[0, 3, 1, 4] + 3\,I[0, 3, 1, 5] -
    \,I[0, 3, 1, 6] - 
   3 \big(I[0, 3, 2, 3] - 2 \,I[0, 3, 2, 4] + I[0, 3, 2, 5]- 
      I[0, 3, 3, 3]  \nonumber\\
      &+ I[0, 3, 3, 4]\big)- I[0, 3, 4, 3]\Big)\Bigg]\Bigg\},
\end{align}
where, $m_c$ is the mass of the c quark, $e_q$ is the electric charge of the corresponding quark,
$\langle \bar qq \rangle$ and $\langle g_s^2 G^2\rangle$ are quark and gluon condensates, respectively.

The functions~$I[n,m,l,k]$, $I_1[\mathcal{A}]$,~$I_2[\mathcal{A}]$,~$I_3[\mathcal{A}]$,~$I_4[\mathcal{A}]$,
~$I_5[\mathcal{A}]$, and ~$I_6[\mathcal{A}]$ are
defined as:
\begin{align}
 I[n,m,l,k]&= \int_{4 m_c^2}^{s_0} ds \int_{0}^1 dt \int_{0}^1 dw~ e^{-s/M^2}~
 s^n\,(s-4\,m_c^2)^m\,t^l\,w^k,\nonumber\\
 I_1[\mathcal{A}]&=\int D_{\alpha_i} \int_0^1 dv~ \mathcal{A}(\alpha_{\bar q},\alpha_q,\alpha_g)
 \delta'(\alpha_ q +\bar v \alpha_g-u_0),\nonumber\\
  I_2[\mathcal{A}]&=\int D_{\alpha_i} \int_0^1 dv~ \mathcal{A}(\alpha_{\bar q},\alpha_q,\alpha_g)
 \delta'(\alpha_{\bar q}+ v \alpha_g-u_0),\nonumber\\
   I_3[\mathcal{A}]&=\int D_{\alpha_i} \int_0^1 dv~ \mathcal{A}(\alpha_{\bar q},\alpha_q,\alpha_g)
 \delta(\alpha_ q +\bar v \alpha_g-u_0),\nonumber\\
   I_4[\mathcal{A}]&=\int D_{\alpha_i} \int_0^1 dv~ \mathcal{A}(\alpha_{\bar q},\alpha_q,\alpha_g)
 \delta(\alpha_{\bar q}+ v \alpha_g-u_0),\nonumber\\
   I_5[\mathcal{A}]&=\int_0^1 du~ A(u)\delta'(u-u_0),\nonumber\\
 I_6[\mathcal{A}]&=\int_0^1 du~ A(u).\nonumber
 \end{align}

\bibliography{refs}
\end{document}